# Usage of Optimal Restructuring Plan in Detection of Code Smells

T.Pandiyavathi

*Student, M.E(SWE), Anna University*

*TamilNadu, India*

*Abstract*- **To remain useful for their users, software systems need to continuously enhance and extend their functionality. Nevertheless, in many object-oriented applications, features are not represented explicitly. The lack of modularization is known to make application features difficult to locate, to comprehend and to modify in isolation from one another. In our work, we implement restructuring using Featureous plug-in where we can change the classes between packages. In the previous literature works, lot of complexities arises while using the tool due to dependencies. Changes to the code without knowing the root cause of the problem leads to further production of new errors. So, we aim in finding the restructuring candidates which have to be rearranged, thereby applying changes to the code on those parts using the tool helps in ordered arrangement of the source code. Applying modularization to the restructuring candidates will lead to decrease in the human effort as well as tool effort in restructuring. Unwanted evolution of new errors will be eliminated**

*Keywords*- **Refactoring, Restructuring, Bad smells**

## I. INTRODUCTION

The design of software systems can exhibit several problems which can be either due to inefficient analysis and design during the initial construction of the software or sometimes may be due to software ageing since software quality degenerates with time. According to Fowler design problems appear as "bad smells" at code or design level and the process of removing them is called refactoring where the software structure is improved without any modification in the behaviour. Restructuring of internal structure of object oriented software along with refactoring is also essential to improve the quality while the software's external behaviour remains unchanged. In previous works, resolution causes ripple effect. Due to unordered sequencing it causes contradictory results which in turn increases human effort and further leads to re-refactoring. Another problem that exists in the previous work is that, restructuring is performed using two or more different tools working in various platforms and resolved separately which also brings increased complexity to the source code. Maximizing the effort-efficiency of feature-oriented changes is of crucial importance. According to several authors, changes concerned with extending, enhancing and correcting application features can amount to 75% of overall cost during software evolution.

## II. RESEARCH WORK

The existing approaches to remodularization generally differ from others with respect to the degree of automation, the applied remodularization criteria and the technical mechanisms of applying these criteria to an existing codebase. The number of existing manual approach greatly exceeds the number of the automated ones during comparison. But, it appears that a significant portion of the existing non-feature-oriented automated approaches could in principle be adapted to using feature-oriented remodularization criteria.

### A. Manual Approaches

Demeyer et al. [6] proposed an idea for reengineering from the ideas learnt from the existing software systems. They provide a holistic set of reengineering patterns that address topics ranging from understanding, redesigning and refactoring a legacy codebase. Thereby placing those code bases in the whole process within an organizational and methodological context is done. Furthermore, they argue for the importance of establishing a culture of continuous reengineering within an organization to enhance flexibility and maintainability of software systems. Demeyer et al. support their proposals with different kinds of applications taken from their industrial experiences.

Mehta and Heineman [14] discussed their experiences from a real-world manual remodularization of an object-oriented application to a feature-oriented decomposition. They present a method for locating and refactoring features into fine-grained reusable components. In a presented case study, feature location is performed by test-driven gathering of execution traces. Features are then manually analyzed, and manually refactored into components that follow a proposed component model. Mehta and Heineman conclude that the performed remodularization improved the maintainability and reusability of features.

The approach of Prehofer was extended by Liu et al. [11], who proposed the feature oriented refactoring (FOR) approach to restructuring legacy applications to feature based decompositions. It aims at achieving a complete separation features in source code, as a basis for creating feature-oriented software product-lines. This is done by the means of base modules that composes of classes and introductions with derivative modules, which contain manually extracted feature-specific fragments of original methods. The approach encompasses an algebraic foundation, a tool 3and a refactoring methodology.

Murphy et al. [12] explored the tradeoffs between three policies of splitting tangled features. By manually separating a set of features at different levels of granularity, he observed a limited potential for tangling reduction of the lightweight approach. In the cases of AspectJ and Hyper/J features, he discovered that using these mechanisms makes code fragments which are difficult to understand in isolation from the others. Aspect-oriented techniques were found to be sensitive to the order of composition, which resulted in coupling of features to one another.

*B. Automated and Semi-Automated Approaches*

Tzerpos and Holt [20] proposed the ACDC algorithm aimed at clustering source code files for aiding software comprehension. To support software comprehension, they equipped their approach with pattern-driven clustering mechanisms, whose aim was to produce decompositions containing well-named, structurally familiar and reasonably sized modules. To achieve this goal, Tzerpos and Holt formalize seven subsystem patterns commonly occurring in manual decompositions of applications. The algorithm was applied to two large-scale software systems to demonstrate that the focus on comprehension does not have a negative effect on the resulting decompositions. To this end, a comparison of ACDC against authoritative decompositions created by developers was performed.

Mancoridis et al. [16] proposed to use cohesion and coupling metrics as the criteria for optimizing the allocation of classes to modules. The metrics were combined into a single objective function to create one measure for evaluating the allocation quality. Mancoridis et al. report on calculations by using this objective function in conjunction with three optimization mechanisms: hill climbing, simulated annealing and genetic remodularization algorithms. The objective function is evaluated based on a simple dependency graph that represents source code files and dependencies among them. The approach was implemented as the clustering tool.

Bauer and Trifu [4] remodularized software using a combination of clustering and pattern-matching techniques. Pattern matching was used to identify structural patterns at the level of architecture. The outputs were then used as an input to a coupling-driven clustering algorithm. To form a single objective function out of six coupling metrics, a weighted sum of the metrics' values was used. The authors recognized that further studies are needed to determine the optimal weights for each coupling metric. The approach was applied to remodularizing the Java AWT library and demonstrated to be superior to a clustering approach oblivious to structural patterns.

O'Keeffe and O'Cinneide [17] defined an approach to optimizing a sequence of refactorings applied to object- oriented source code to improve its adherence to QMOOD which is a better quality model. This is done by formulating the task as a search problem in the space of alternative designs and by automatically solving it using hill climbing and simulated annealing techniques. The search process is driven by a weighted sum of eleven metrics of the QMOOD. In application to two case studies, the approach was demonstrated to successfully improve the quality properties defined by QMOOD.

### 3. MODULES

*A. Data Collection*

Select Sample code to restructure whose quality attributes have to be efficient. Parse the given source code. Simple parsing is done using online tool

*B. Refactoring*

Identify the complexity in the source code using traditional metric. Metrics tool is used for calculating the complexity in the sample code. Complexity arises in reducing the level of complexity in the refactored code. Two main measures used are McCabe Cyclomatic complexity and Source Lines of Code

PMD: Detects dead code, duplicate code, long method, long parameter list

Checkstyle: Detects long method and feature envy

As far as refactoring is concerned we learn about using genetic algorithm for refactoring the code with refactoring methods as well as sequencing of the code smells and refactoring methods is also done. Genetic algorithm implements the partially mapped crossover due to conflicts which occurred while using the 1-point crossover and 2-point crossover techniques.

Refactoring steps

1. Perform pair wise analysis among each selected code smells.
2. Draw directed graph based on the analysis made above
3. Apply topological sorting to obtain ordered code smells.
4. Generate detection rules using combinations of metrics and thresholds
5. Collect refactorings methods to be processed after the defect detection
6. Generate/ frame list of possible refactoring methods like pull up, move method, extract method
7. Apply natural evolution techniques like genetic algorithm with input as the outcomes of steps 4,5,6
8. Perform crossover and mutation along with the elitism property in the above algorithm
9. Obtain Optimal solution with sequenced refactoring plans

*C. Check Cohesion Levels & Determine Restructuring Candidates*

Java code is implemented to find the comment line and code to comment ratio to find the complexity and the metric LCOM calculated using metrics. Calibration of the calculated values with "rule of 30" and Restructuring candidates (RC) is generated

Cohesion levels should be checked and the design is made for calculation and for categorizing the class with cohesion levels as (L, M, H) and the calculation is tried for initial simple java program Calculator.java

Fig. 1 Statistics Collected From Code Analyser

*D. Restructuring*

Design should correctly implement all the needs on both functional and non-functional requirements and also enable efficient code production & division of work among team members and should be minimal as simple as possible addressing all requirements. The approach should be clear and feasible to read which allows intellectual control over a project and manage qualities like complexity, maintainable and extensible. Here we aim in trying both refactoring and restructuring techniques in our approach to obtain a quality code for easy maintenance and understandability of the source code. Two main features namely, scattering and tangling values can be monitored and based on those values restructuring can be done. Here we follow 2 methods

1. Finding the candidates which needs to be restructured using some defined criteria's and rule of 30

2. Analyzing the dependencies between packages, classes, and methods using the plug-in and refactor them manually

*Enabling The Featureous Plugin*

i) Select the project

ii) Set the project as the main project from the properties menu

iii) If the packages are successfully imported, then we will get a "trace project" button in our tool bar

iv) Click on the trace project button

v) A xml file will be created in the project folder. Open the file and change the root to the project package name

vi) Save it and close the file

vii) Clean, build and trace the project again. Traces will be created, but we need to annotate the code

viii) Annotate the method names as @FeatureEntryPoint("sample")

ix) Clean, build and trace again and give appropriate values. Traces will be successfully created and updated into the folder

x) Open windows-> others-> Featureous

Modularize the code/Based on color scheme find the dependency and do restructuring

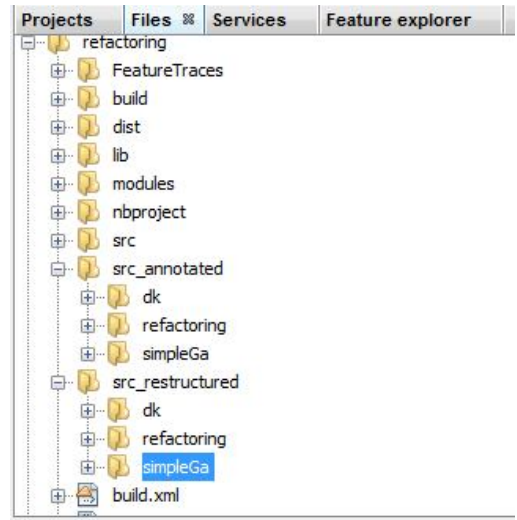

Fig. 2 Restructured Source Code

*Values of Metrics without Refactoring/Restructuring*

1) Complexity measured in average for the source code used is 7.048
2) Total scattering of features among packages (FSCA) ranges approximately 0.247
3) Total tangling of features in packages (FTANG) ranges approximately 0.205
4) Average package cohesion (PCOM) ranges approximately 0.648

5) Total coupling of packages (PCOUP) ranges approximately 0 in our sample code

*Values of metrics with Refactoring/ Restructuring*

1) Complexity measured in average for the source code used is 4.59
2) Total scattering of features among packages (FSCA) ranges approximately 0.208
3) Total tangling of features in packages (FTANG) ranges approximately 0.202
4) Average package cohesion (PCOM) ranges approximately 0.421
5) Total coupling of packages (PCOUP) ranges approximately 0 in our sample code

## 4. CONCLUSION

*A. Contribution*

Since we can change the classes between packages, restructuring brings additional injection of errors. To avoid this, timer is set before restructuring for the process of changing classes between packages. Even though this may not stop the havoc caused in the source code, it may still minimize the effect.

Applying modularization to the restructuring candidates will lead to decrease in the human effort as well as tool effort in restructuring. Unwanted evolution of new errors will be eliminated

*B. Limitation*

Annotation of the restructuring candidates will take more time. Only with experienced people, annotation work can be done in time efficient manner. But this not a serious issue to be considered since time taken in manual refactoring takes enormous time. Another drawback that we have encountered is space requirements. The method without refactoring and restructuring has no need of much space requirements. But for using restructuring and applying it to the program, some plugins and tools should be installed and it needs more space as compared to the previous works. Since we don't have any problem in providing the extra space requirements this is not considered as major drawback of the system

*C. Future Enhancement*

Efficient way of automated annotation to the source code can be implemented by making changes to the Featureous plugin. Even though automated annotation is present in the current tool, it leads to creation of many errors as well as not all the annotated source codes are executed without exceptions. So effective way should be applied before entering the restructuring step, which we perform in the Featureous plugin.

## V. RESULTS

We observe reduction of 11% and 7% of FSCA, FTANG in the source code after restructuring. We also observe acceptable level of increase and decrease in the cohesion and coupling rates

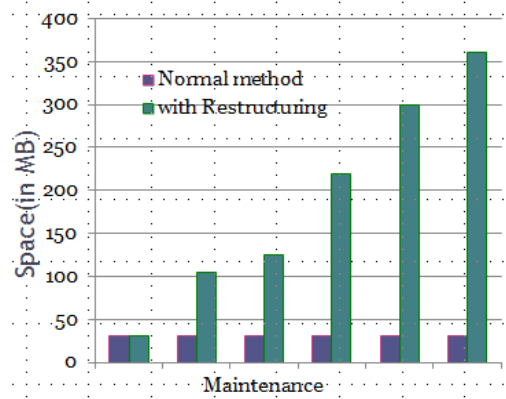

Fig. 3 Space Vs Time

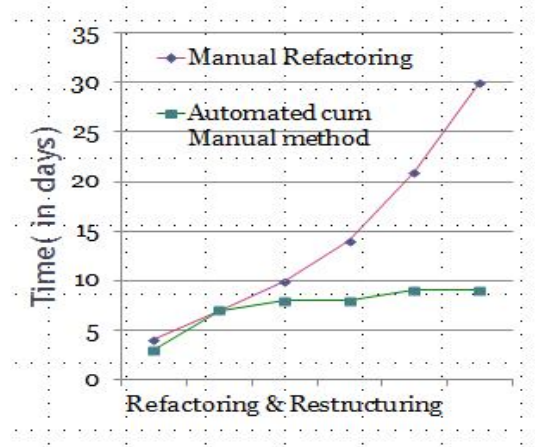

Fig. 4 Time Vs Restructuring

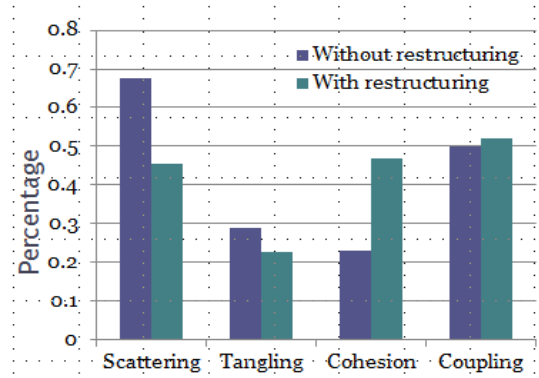

Fig. 5 Metrics Vs Restructuring